\documentclass[twocolumn]{aastex62}

\usepackage{graphicx}
\usepackage{hyperref}
\usepackage{amssymb,amsmath}

\newcommand {\xmm} {\textit{XMM-Newton}}
\newcommand {\chandra} {\textit{Chandra}}
\newcommand {\suzaku} {\textit{Suzaku}}

\newcommand {\hess} {H.E.S.S.}

\newcommand {\spitzer} {\textit{Spitzer}}

\newcommand{\wise} {\emph{WISE}}

\newcommand{\gray}{$\gamma$-ray\,}

\def\lea{\mathrel{<\kern-1.0em\lower0.9ex\hbox{$\sim$}}}
\def\gea{\mathrel{>\kern-1.0em\lower0.9ex\hbox{$\sim$}}}
\newcommand{\lta}{{\>\rlap{\raise2pt\hbox{$<$}}\lower3pt\hbox{$\sim$}\>}}
\newcommand{\gta}{{\>\rlap{\raise2pt\hbox{$>$}}\lower3pt\hbox{$\sim$}\>}}

\def \rsun {\ifmmode$R$_{\odot}\else R$_{\odot}$}

\newcommand {\hone} {H~{\small I}}

\newcommand {\hmol} {H$_2$}

\begin{document}

\title{Runaway O-star Bow Shocks as Particle Accelerators? The Case of AE Aur revisited}

\author[0000-0002-9282-5207]{Blagoy Rangelov}
\affil{Department of Physics, Texas State University, 601 University Drive, San Marcos, TX 78666, USA}

\author[0000-0002-4304-9846]{Thierry Montmerle}
\affil{Sorbonne Universit{\'e}, CNRS, UMR~7095, Institut d'Astrophysique de Paris, 98bis, Bd Arago, 75014 Paris, France}

\author[0000-0002-8433-9663]{S.~R. Federman}
\affil{Department of Physics \& Astronomy, The University of Toledo, 2801 West Bancroft Street, Toledo, OH 43606, USA}

\author[0000-0002-4304-9846]{Patrick Boiss{\'e}}
\affil{Sorbonne Universit{\'e}, CNRS, UMR~7095, Institut d'Astrophysique de Paris, 98bis, Bd Arago, 75014 Paris, France}

\author{Stefano Gabici}
\affil{APC, Universit{\'e} Paris Diderot, CNRS/IN2P3, CEA/IRFU, Observatoire de Paris, Sorbonne Paris Cit{\'e}, F-75013 Paris, France}

\shorttitle{Chandra observations of AE Aur}
\shortauthors{Rangelov et al.}

\begin{abstract}

We present results of our \chandra/ACIS observations of the field centered on the fast, runaway O star AE Aur and its bow shock. Previous \xmm\ observations revealed an X-ray ``blob'' near the IR arc tracing the bow shock, possibly a nonthermal source consistent with models of Inverse Compton scattering of dust IR photons by electrons accelerated at the shock. The new, subarcsecond resolution \chandra\ data, while confirming the presence of the \xmm\ source, clearly indicate that the latter is neither extended nor coincident with the IR arc and strongly suggest it is a background AGN. Motivated by results published for the bow shock of BD+43$^{\circ}$3654, we extended our study to the radio domain, by analyzing archival EVLA data. We find no radio emission from the AE Aur bow shock either. The corresponding upper limits for the absorbed (unabsorbed) X-ray flux of $5.9(7.8)\times10^{-15}$\,erg\,cm$^{-2}$\,s$^{-1}$ (3$\sigma$) and, in the radio range, of 2\,mJy (1.4\,GHz), and 0.4\,mJy (5.0\,GHz), are used to put constraints on model predictions for particle acceleration within the bow shock. In the ``classical'' framework of Diffusive Shock Acceleration, we find that the predicted X-ray and radio emission by the bow shock is at least two orders of magnitude below the current upper limits, consistent with the systematic non-detections of up to 60 stellar bow shocks. The only exception so far remains that of BD+43$^{\circ}$3654, probably the result of its very large mass-loss rate among runaway O stars.

\end{abstract}


\section{Introduction}

The origin of cosmic rays (CRs), and more generally the investigation of the physical mechanisms able to accelerate particles (electrons and nuclei) up to energies as high as $10^{21}$ eV, remains one of the most fascinating problems in astrophysics. On a galactic scale, the origin of CRs is widely attributed to their acceleration by supernova shock waves, seen as supernova remnants (SNRs) expanding supersonically in the interstellar medium (ISM). Nuclei are able to travel for millions of years throughout the Galaxy and the Galactic halo (e.g., \citealt{Gaisser2016}), losing their energy along the way by spallation collisions in the non-relativistic regime. This is deduced from various secondary/primary abundance ratios at Earth, and, above $\sim 1$ GeV/n,  by $\pi^0$-decay induced $\gamma$-ray emission, clearly seen in particular along the Galactic plane (e.g., \citealt{Strong2007}). On the other hand, electrons generally lose their energy on much smaller distance scales by radiative emission (bremsstrahlung, synchrotron), detectable with radio telescopes usually in the cm range, or by collisions with ambient photons (Inverse Compton; IC), boosting their energies up to X-rays or even $\gamma$-rays. 

So, from the observational point of view, one must distinguish between ``direct'' clues to investigate the acceleration processes (e.g., radio observations), and ``indirect'' ones, i.e., those necessitating a target to be revealed such as molecular clouds for relativistic protons \citep{Abdo2010,Acciari2009,Katsuta2012,Katagiri2016,Tavani2010}, or radiation from interstellar dust for electrons.

In this context, by far the most popular acceleration mechanism is the so-called ``Diffusive Shock Acceleration'' (DSA) (e.g., \citealt{Drury1983}). This mechanism, in various forms, has been successfully tested (at the cost of adjusting some parameters) in a variety of environments, either isolated SNR (for instance in relation with historical supernovae, e.g., SN1006), middle-aged SNR in regions of star formation, or even colliding winds in massive binary systems \citep{Petruk2017,Cardillo2017,Uchiyama2010,Gabici2009,Malkov2011,Torres2010}.

In this paper, we want to investigate further another situation where cosmic-ray acceleration (here, electrons) could potentially take place: that of high-velocity, massive stars (called ``runaways'') traveling supersonically through a dense enough ISM. In such conditions, these stars are preceded by a more or less paraboloid-shaped ``bow shock'', resulting from the collision between the stellar wind and the ambient ISM (e.g., \citealt{BM1988,Peri2012}). In fact, direct detection of electrons accelerated by the bow shock of a very massive star, by way of their synchrotron radio emission, was first obtained by \citet{Benaglia2010} for BD+43$^{\circ}$3654 (O4f).

\begin{figure}
\includegraphics[scale=0.105]{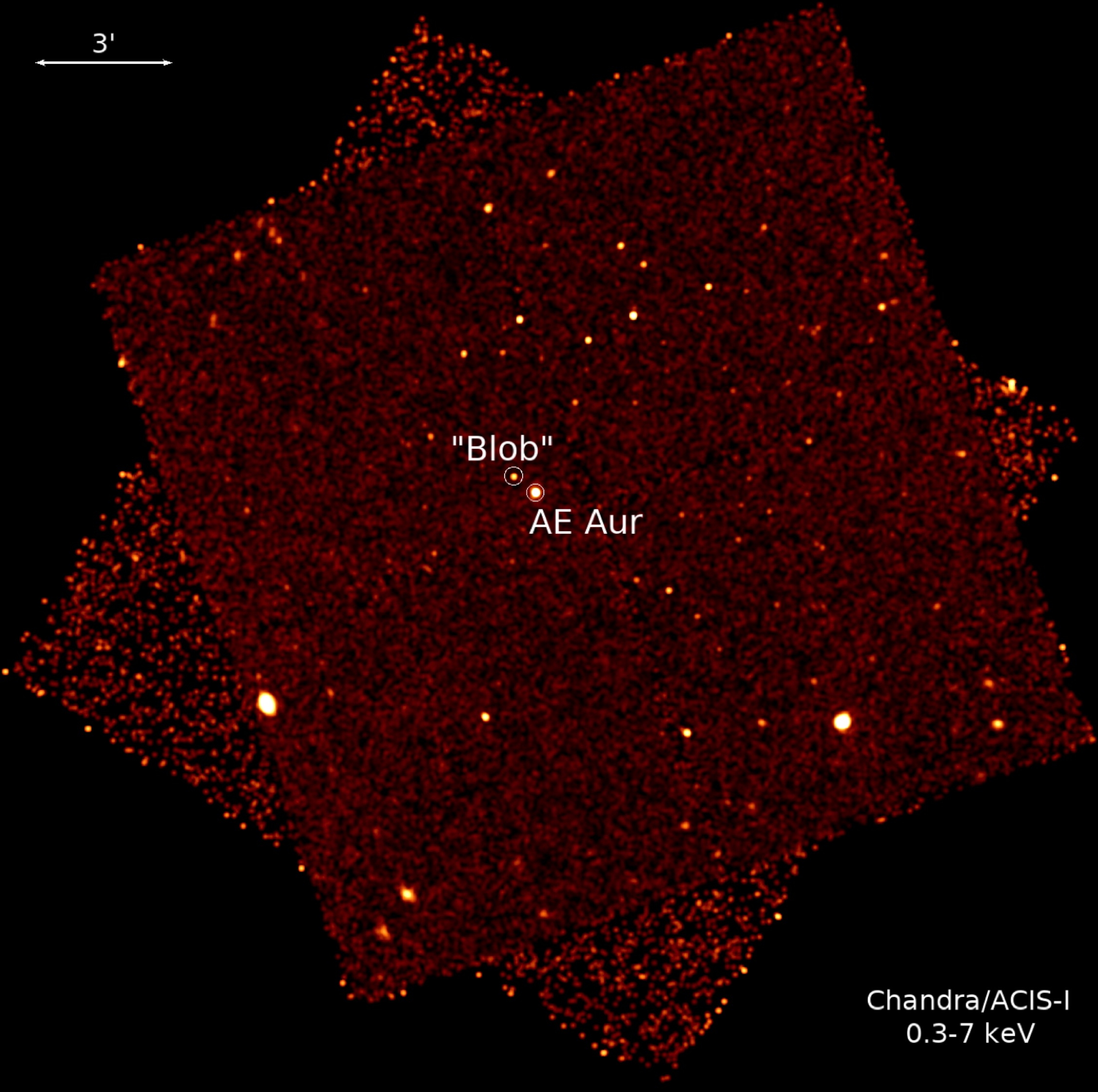}
\caption{Exposure corrected \chandra\ ACIS-I merged image of AE Aur in the 0.3--7\,keV energy band smoothed with a $5''$ kernel. The two brightest sources are AGNs (see text); the source at the center is our target, AE Aur, and the source to the north-east corresponds to the XMM ``LS blob'' detected by \citet{Lopez-Santiago2012}. 
North is up and East is to the left.}
\label{chandra_merged}
\end{figure}

This discovery prompted \citet{VR2012} to show that, under certain conditions, electrons similarly accelerated at the bow shock could be revealed (albeit indirectly) by boosting IR photons from the shock-compressed dust to X-ray energies, up to an observable level. Following this work, \citet{Lopez-Santiago2012}, hereafter LS12, used \emph{XMM-Newton} archival data to study one of the fastest known runaway stars associated with a bow shock, the late-O (O9.5V) star AE Aur. This star is travelling at $v_\star = 150$\,km\,s$^{-1}$ in a moderately dense ISM \citep{Tetzlaff2011} and has been ejected from the Orion nebula a few millions years ago \citep{Hoogerwerf2001}; it is located at a distance of 530 pc and its mass-loss rate is $\dot{M} = 2 \times 10^{-7} M_\odot$\,yr$^{-1}$. A $\sim$40~ks exposure (bin size $4''$) revealed a seemingly extended X-ray ``blob'' spatially correlated with an arc-shaped mid-IR dust feature seen with the {\it Wide-field Infrared Survey Explorer} (\emph{WISE}; the spatial resolution is 6\farcs5 at 12~$\mu$m: \citealt{Cutri2012}), although not with its apex. Their analysis of the X-ray spectrum (1-8 keV) favored non-thermal emission (photon index $\Gamma \sim -2.6$). Using the model by \citet{VR2012}, this emission could be explained by a power-law (PL) electron injection spectrum. Adjusting some free parameters (like the acceleration efficiency), the resulting photon spectrum was found to peak in the soft X-ray band and account for the observed X-ray luminosity.

However, \citep{Toala2017}, based on their own analysis of \emph{XMM-Newton} archival data, and of the higher angular resolution of \emph{Spitzer} images obtained by \citet{France2007} compared to \wise\ (IRAC resolution of 2 arcsec at 8 $\mu$m), showed that the ``LS blob'' of X-ray emission near AE Aur is not spatially coincident with the bow shock. In addition, their analysis of five other runaway stars yielded only upper limits to their bow-shock X-ray fluxes.

\begin{figure}
\begin{center}
\includegraphics[scale=0.37]{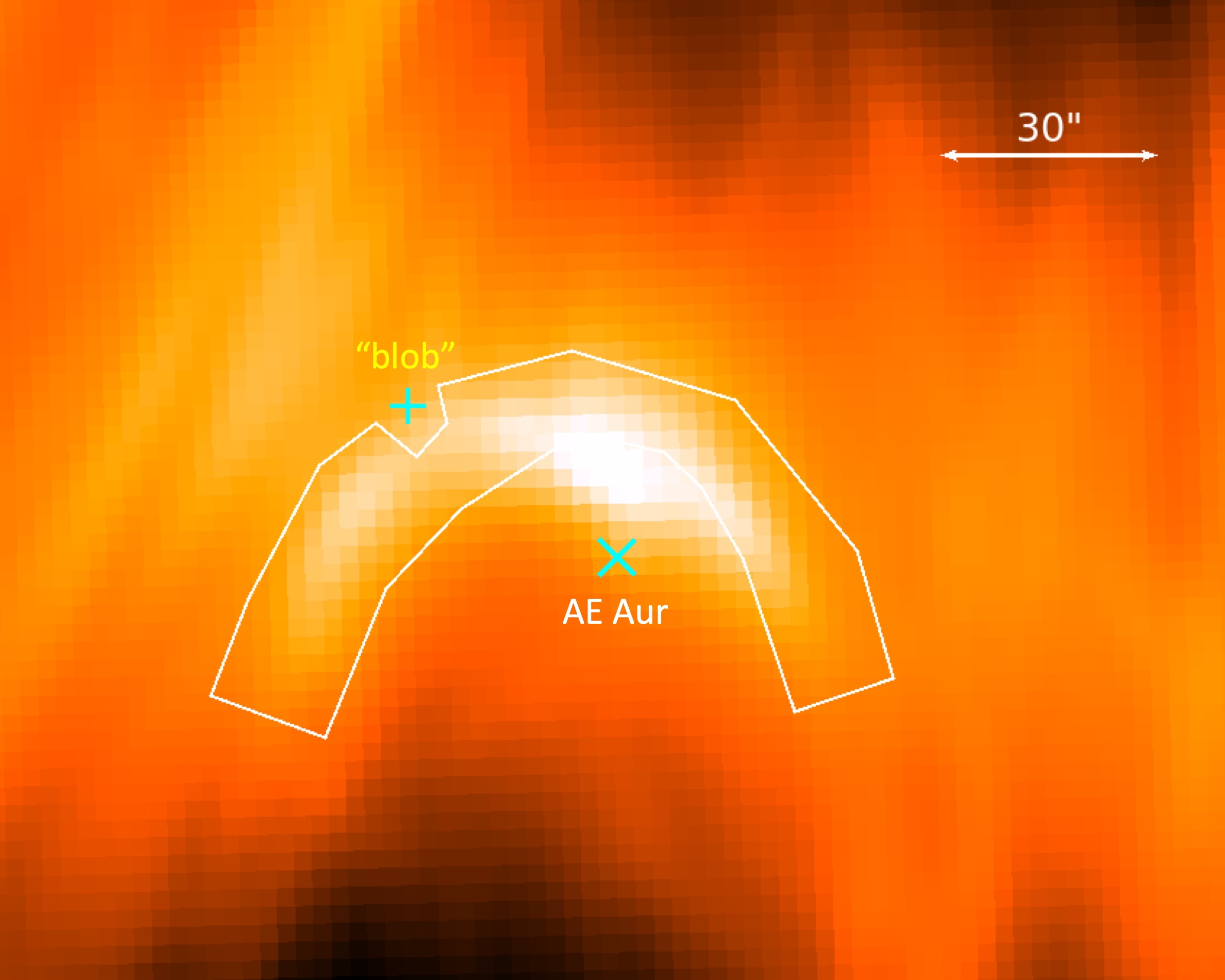}
\caption{AE Aur and its bow-shock in {\it Spitzer} 24\,$\mu$m band (this band is preferred to other shorter wavelength ones because it is less contaminated by emission from AE Aur and less affected by saturation). The white region shows the bow-shock extraction region used for X-ray analysis (see text for details). The ``LS blob'' and AE Aur are shown with $+$  and $\times$ signs, respectively. The contour avoids the ``blob" region to exclude its X-ray flux.}
\label{24mu}
\end{center}
\end{figure}

In view of these conflicting results,  we obtained several {\it Chandra}/ACIS images ($17' \times 17'$ FOV) having a much better angular resolution ($\sim$$0\farcs5$ on-axis)  than that of \xmm, to map out the area around AE Aur in detail and its bow shock. The resulting merged image is shown in Figure~\ref{chandra_merged}. 

The outline of the paper is as follows. We first present an analysis of our observations, including other sources in the field to help characterize the nature of the ``LS blob'' (Section \ref{obs}). In Section~\ref{results}, we find very clear evidence that the \xmm\ ``LS blob'' is actually a faint point source with no counterpart at other wavelengths and unrelated to the bow shock, but having a definitely hard, non-thermal spectrum. Further analysis of the global \chandra\ image suggests that the ``LS blob'' is likely a background AGN. {Moreover, we find no diffuse emission from the area delineated by the IR arc, neither in the \chandra\ X-ray image, nor in archival EVLA radio data. In Section \ref{discussion}, we put AE Aur in the context of other well-studied runaway stars with bow shocks and discuss the resulting constraints on theoretical models of particle acceleration by stellar bow shocks (Section 5). Finally, we summarize our main conclusions in the last Section (\S\ref{conclusions}). A future, standalone paper will be dedicated to the X-ray properties of AE Aur as a mass-losing, late-type O star.

\section{Observations and Data Reduction}
\label{obs}

\subsection{Chandra X-ray Observatory}

We carried out a campaign to observe the region of AE Aur with \chandra. The program (PI: Rangelov) was split into five observations totaling 140.53\,ks exposure over the period of two months. The first data set was taken on 2016 December 16 (ObsID 19943; 14.88\,ks), followed by 2016 December 17 (ObsID 19445; 44.49\,ks), 2017 January 3 (ObsID 19979; 26.72\,ks), 2017 January 4 (ObsID 19941; 26.72\,ks), and 2017 January 6 (ObsID 19951; 27.72\,ks). All data were taken with the ACIS-I instrument operated in ``VeryFaint'' Timed exposure mode. We processed the data using the {\it Chandra} Interactive Analysis of Observations (CIAO\footnote{\\{http://cxc.cfa.harvard.edu/ciao/}}) software version 4.8 and {\it Chandra} Calibration Database version 4.7.2. The data have been restricted to the energy range between 0.3 and 7\,keV and filtered in three energy bands, 0.3--1.2\,keV (soft), 1.2--2\,keV (medium), and 2--7\,keV (hard). We used the CIAO's Mexican-hat wavelet source detection routine \texttt{wavdetect} \citep{Freeman2002} to create source lists. Wavelets of 1.4, 2, 4, 8, and 16 pixels and a detection threshold of $10^{-6}$ were used, which typically results in one spurious detection per million pixels. In order to find fainter point sources, all five datasets were merged\footnote{Standard CIAO procedures found at \url{http://cxc.harvard.edu/ciao/threads/wavdetect_merged/} were followed to merge the data. We used an exposure-time-weighted average PSF map in the calculation of the merged PSF.} prior to running \texttt{wavdetect}. We detected a total of 114 X-ray sources in the merged data. The \texttt{srcflux} CIAO tool was then run individually on each observation (using the coordinates found by \texttt{wavdetect}).  

In the following, we adopt the designation CAX-\textit{nn} for these sources, where ``C'' stands for \chandra\ and ``A'' for AE~Aur; \textit{nn} is the rank when sources are ordered by increasing right ascension. An analysis of sources in the whole \chandra\ image is presented below (\S 3.5).

\subsection{XMM-Newton Telescope}

We used archival data from \xmm\ (PI Damiani, ObsID 0206360101) with total exposure time of 58.9\,ks. This is the same dataset as that analyzed by LS12. We re-processed the data ourselves using the \xmm\ Science Analysis System (SAS\footnote{\url{https://www.cosmos.esa.int/web/xmm-newton/sas}}) software version 17.0.0 and followed the standard source extraction procedures from the SAS Data Analysis Threads\footnote{\url{https://www.cosmos.esa.int/web/xmm-newton/sas-threads}} to extract spectra from MOS1, MOS2 and PN (see Section\,\ref{results} for details).

\subsection{Multi-Wavelength Data Analysis}
\label{mw}

We searched multi-wavelength (MW) catalogs for potential counterparts to all X-ray sources and compiled a set of MW parameters for each of the sources with counterparts. We collected optical measurements from the USNO-B \citep{Monet2003} catalog, the near-infrared (NIR) from the \textit{Two Micron All-Sky Survey} (2MASS; \citealt{Skrutskie2006}), and the IR from \emph{WISE} \citep{Cutri2012}. MW sources were considered potential counterparts if they were located within the error radius of each X-ray source. We used a 95\% confidence level positional uncertainty of $1\farcs5$ typical for a $5'$ off-axis X-ray source with 15 counts (see Equation 12 in \citealt{Kim2007}). The X-ray properties and magnitudes for the MW counterparts (up to 11 MW parameters) are then used to classify these sources with our machine-learning pipeline (see the Appendix in \citealt{Hare2016} for details). 

To help understand the large-scale topography of the ISM in the direction of our \chandra\ field, Figure~\ref{24mu} shows the 24\,$\mu$m \emph{Spitzer/MIPS} image covering a $\sim 2^\circ \times 2^\circ$ area centered on AE Aur within the merged \chandra\ image (Figure~\ref{chandra_merged}).
AE~Aur corresponds to CAX-72 while the \chandra\ counterpart to the \xmm\ ``LS blob'' is CAX-76.

\section{Bow shock Results and Analysis}
\label{results}

Before discussing the properties of CAX-76 and presenting a global analysis of other sources detected in our \chandra\ image, we summarize the information available on the interstellar medium in the field. This material induces significant absorption which must be taken into account when fitting \chandra\ spectra, especially at low energies.

\subsection{The interstellar medium in the field of AE Aur}

\hone\ and \hmol\ column densities in front of AE Aur were inferred from UV spectra and amounts to $N$(\hone) = 2.7$\times10^{21}$\,cm$^{-2}$ and $N$(\hmol) = 6.4$\times10^{20}$\,cm$^{-2}$ \citep{Boisse2005}, leading to a total H column density, $N_{\rm H}$ = $N$(\hone) + $2 \times N$(\hmol) = $4.0\times10^{21}$\,cm$^{-2}$, appropriate for computing AE~Aur's X-ray absorption.

\begin{figure}
\begin{center}
\includegraphics[scale=0.37]{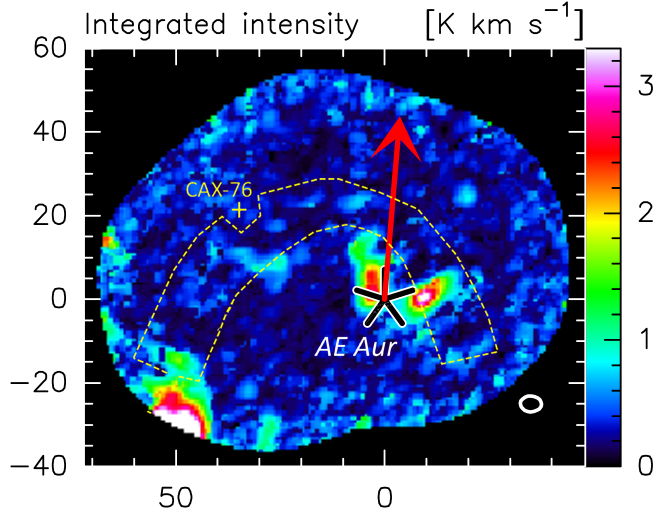}
\caption{
CO mapping of a $\sim 1.5' \times 1.5'$ area covering AE Aur and its bow shock (Pierre Gratier, private communication). The yellow dotted line encompasses the IR bow shock, as seen in the \emph{Spitzer} image (Figure~\ref{24mu}), excising the ``blob'' X-ray source CAX-76. The red arrow indicates the proper motion of the star, which is actually located behind the CO clumps, explaining its relatively high extinction (see text for details.)
}
\label{CO}
\end{center}
\end{figure}

Figure~\ref{CO} shows the $\sim 1.5' \times 1.5'$ CO mapping at $4''$ resolution by \citet{Gratier2014}, revealing two tiny molecular clumps along the AE Aur line of sight. However, no CO emission has been detected toward CAX-76 and the area delineated by the IR arc is also essentially devoid of CO emission. Thus, for a source lying near AE~Aur in space, the appropriate $N_{\rm H}$ value should be around 2.7$\times10^{21}$\,cm$^{-2}$ because there is little molecular material in the vicinity. 

For extragalactic sources, the total Galactic $N$(\hone) value in this direction should be adopted, $N$(\hone) = 5.8$\times10^{21}$\,cm$^{-2}$, as provided by the HI4PI survey \citep{Bekhti2016}. This value is in fact a lower limit since additional absorption, internal to the sources, might be present. The HI4PI survey indicates that spatial variations across the \chandra\ field are very limited ($N$(\hone) ranges from 5.6 up to 6.0$\times10^{21}$\,cm$^{-2}$), thus the above $N$(\hone) value should hold throughout the field. CO mapping at $22''$ resolution over a significant portion of the \chandra\ field (Gratier et al., in preparation) indicates that molecular gas is present in the form of dense clumps but that their surface covering factor is no larger than a few percent; thus, for most background sources, X-ray absorption can be computed on the basis of \hone\ data alone (i.e., $N_{\rm H} \approx N$(\hone)).

\subsection{Properties and nature of source CAX-76}

\begin{deluxetable*}{lrrcccc}
\tablecolumns{7}
\tablecaption{X-ray model fit parameters for the ``blob'' source (CAX-76)}
\tablehead{ \colhead{Data} & \colhead{Model} & \colhead{N$_{\rm H}$\tablenotemark{a}} & \colhead{kT\tablenotemark{b}} & \colhead{$\Gamma$\tablenotemark{c}} & \colhead{$\chi^2$} & \colhead{DOF} \\
\hline
\multicolumn{7}{c}{Models with fixed N$_{\rm H}$}
}
\startdata
\chandra\ & APEC & 2.7 & 64 & -- & 24.92 & 18 \\
\chandra\ & PL & 2.7 & -- & $1.1\pm0.3$ & 24.42 & 18 \\
\xmm\ & APEC & 2.7 & $6\pm3$ & -- & 6.05 & 13 \\
\xmm\ & PL & 2.7 & -- & $1.8\pm0.3$ & 6.75 & 13 \\
\chandra+\xmm\ & APEC & 2.7 & 26 & -- & 37.27 & 32 \\
\chandra+\xmm\ & PL & 2.7 & -- & $1.3\pm0.2$ & 37.39 & 32 \\
\cutinhead{Models with free N$_{\rm H}$}
\chandra\ & APEC & 3 & 64 & -- & 24.89 & 17 \\
\chandra\ & PL & $10^{-4}$ & -- & $0.8\pm0.5$ & 24.88 & 17 \\
\xmm\ & APEC & $8\pm4$ & $2.8\pm1.6$ & -- & 4.86 & 12 \\
\xmm\ & PL & $9\pm5$ & -- & 2.6$\pm$0.7 & 4.49 & 12 \\
\chandra+\xmm\ & APEC & 1.9 & 55 & -- & 37.13 & 31 \\
\chandra+\xmm\ & PL & 2 & -- & $1.3\pm0.4$ & 37.29 & 31 \\
\enddata 
\label{fits} 
\tablenotetext{a}{Hydrogen column density in units of $10^{21}$\,cm$^{-2}$ for \texttt{phabs} model.}
\tablenotetext{b}{Temperature for APEC model in units of keV.}
\tablenotetext{c}{Photon index for PL model.}
\tablecomments{Parameters without listed uncertainties indicate models where given parameter reaches the lower/upper parameter boundary during the fit.}
\end{deluxetable*}

We detect a weak source, CAX-76, consistent with the location of the original ``LS blob'' reported by LS12. The source has the following right ascension (RA) and declination (DEC), RA = 5:16:20.45 and DEC = +34:19:05.18. Its offsets with respect to AE~Aur are $\Delta {\rm RA}= 34\farcs3$ and $\Delta {\rm DEC}=30\farcs8$ respectively. Contrary to the findings of LS12, CAX-76 appears to be point-like at the \chandra\ resolution, implying an extent no larger than about 0.5~arcsec. The source has no counterpart in the MW catalogs. We have investigated optical/NIR/IR images (e.g., \emph{DSS}, \emph{2MASS}, \emph{Spitzer}) to see whether its X-ray emission could be attributed to a background, uncatalogued (flaring) star or AGN, but we find no counterpart either. The closest point-like stellar source, 2MASS~05162332+3420290 ({\it JHK} magnitudes 15.863, 15.428 and 15.208, respectively), is $15\farcs5$ away from the CAX-76, too far to be the X-ray source. 

We collected 106 counts for CAX-76 in the total (from all five observations), 140\,ksec \chandra\ observation and used CIAO's task \texttt{combine\_spectra} to combine all five spectra into one. In order to improve the S/N ratio, we also extracted the \xmm\ spectrum in a similar fashion to LS12, using an annular background region around AE Aur, with the inner and outer radii of the annulus placed in such a way to tightly encompass the ``LS blob'' (which was of course excluded from the background extraction). LS12 used only the PN spectrum for their analysis because they report that the MOS spectra are not constraining enough. We have checked that the inclusion of the MOS spectra does not change our conclusions, but we analyze further the \xmm\ PN data only, to maintain a more direct comparison to the LS12 results. 

Due to the limited number of counts at low energies ($\lessapprox1$\,keV), it is difficult to constrain the fit to the X-ray data. Figure~\ref{x-ray_blob} shows the \chandra\ and \xmm\, X-ray spectra. We have attempted to fit the data with absorbed PL ($A(E)=K E^{-\Gamma}$, where $\Gamma$ is the photon index and $K$ is the normalization) and APEC (emission spectrum from collisionally-ionized diffuse gas) models in {\it XSPEC} \citep{Arnaud1996}. For each model and data set, we first fixed $N_{\rm H}= N_{\rm H}{\rm(AE~Aur)}$, then took it as a free fitting parameter. Results are presented in Table~\ref{fits}. We do not provide uncertainties for parameters that reach the lower/upper parameter boundary during the {\it XSPEC} fit. In these cases, the fit provides unrealistic parameters, such as very large kT values for the APEC models. 

Our \xmm-only formal PL fit yields $\Gamma = 1.8\pm0.3$ for $N_{\rm H}$ fixed  to $N_{\rm H}$(AE Aur). This is different (although consistent within 1$\sigma$) from the value reported by LS12, $\Gamma=2.6\pm0.6$. While we tried to reproduce the \xmm\ CAX-76 extraction procedures used by LS12, we may have taken a slightly different background region, which is subject to contamination by photons from AE Aur itself. This affects more heavily the lower energy X-ray photons (AE Aur completely disappears above $\gtrapprox2$\,keV). Altogether, within our analysis we find rather consistent estimates for the indices of the X-ray PL fits (see Table~1): $\Gamma=1.8\pm0.3$ for our \xmm\ reprocessed value, $\Gamma=1.1\pm0.3$ with our \chandra\ data, and $\Gamma=1.3\pm0.2$ when combining our \xmm\ and \chandra\ spectra. Therefore, we concur with LS12 in that there is little doubt that the CAX-76 spectrum, although difficult to fit, is hard and cannot be thermal. Note that, contrary to \xmm, there is no \chandra\ data below 1\,keV. This will affect the spectral fits, especially the derived extinction. As discussed below (Section~\ref{chandra_global}), the analysis of other X-ray sources in the \chandra\ field provides further evidence that CAX-76 is most likely a background AGN. The various values of $\Gamma$ found for CAX-76 can be compared with the ones found for AGNs in the 2$-$10\,keV interval, which lie in the range $\Gamma \sim 1.5-2.5$ (e.g. \citealt{IC2010}, and refs. therein). Within errors, the values are compatible, which is consistent with the suggestion that CAX-76 (the "LS blob") is a background AGN. \\ \\

\begin{figure}
\begin{center}
\includegraphics[scale=0.165]{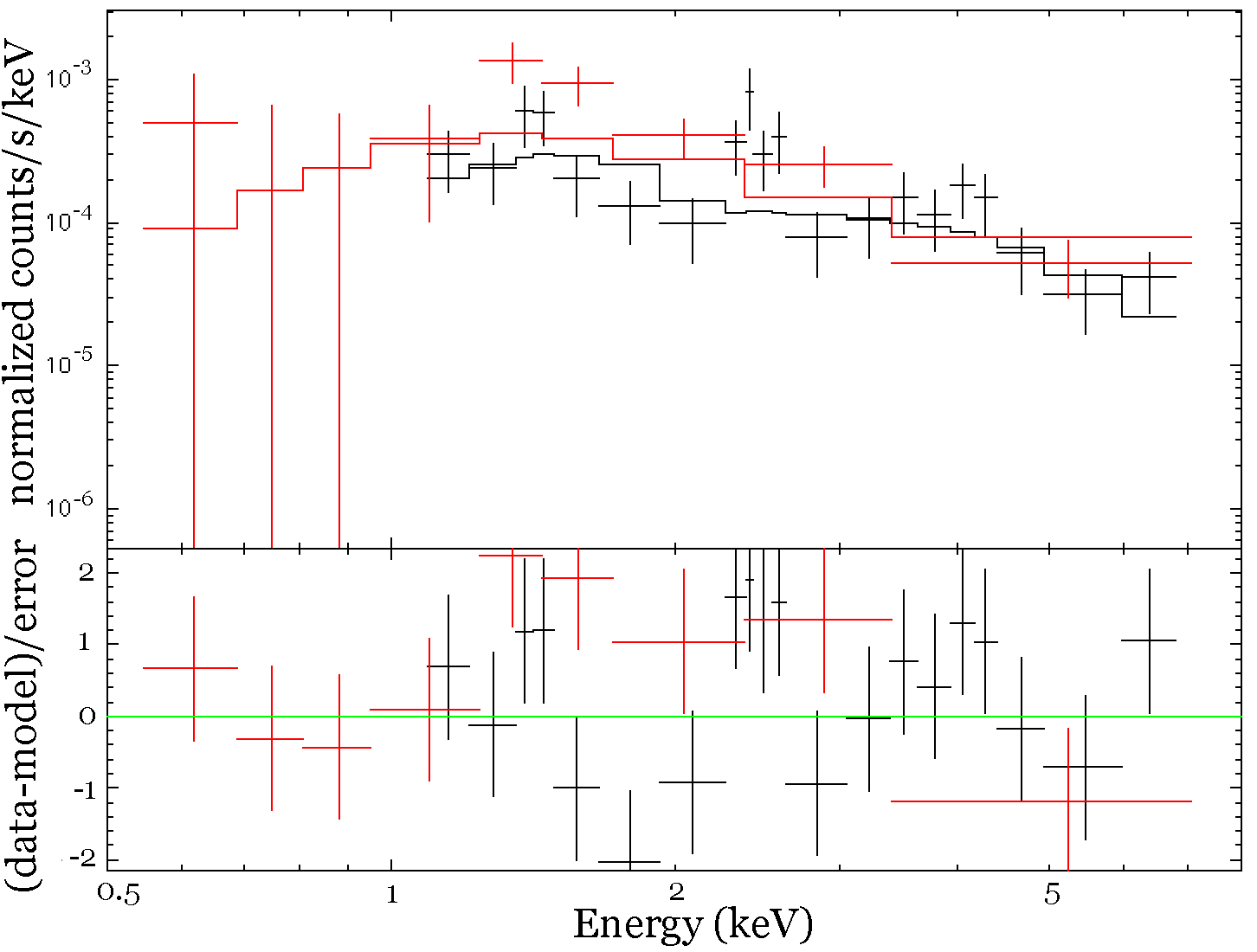}
\caption{\chandra\ and \xmm\, X-ray spectra of CAX-76 shown in red and black, respectively. A simple absorbed PL ($\Gamma=1.3$) model is used for illustration (see text and Table~\ref{fits} for details).}
\label{x-ray_blob}
\end{center}
\end{figure}

\subsection{Zooming in: X-rays from the bow shock?}

We find no evidence of extended X-ray emission in the vicinity of AE Aur associated with its bow shock. Using the exposure corrected X-ray and \spitzer\ 24\,$\mu$m images (which do not show saturation), we created an \textit{ad hoc} extraction region covering the apex of the bow shock (shown in Figure~\ref{24mu} and Figure~\ref{CO}), designed in such a way as to avoid contamination from point sources (specifically CAX-76 and CAX-72) and chip gaps. We measured a count rate of $1.53\times10^{-3}$\,cts\,s$^{-1}$ (0.3--7\, keV band) in this 1771\,arcsec$^2$ extraction region. For comparison, we obtained the upper limit to the bow-shock flux by sampling the background count rates from 12 different regions in the vicinity of AE Aur and by calculating the standard deviation from the mean value. The background regions had the same shape and size as the bow-shock region. The average count rate and standard deviation are $(1.62\pm0.11)\times10^{-3}$\,cts\,s$^{-1}$. We consider the standard deviation as a conservative $1\sigma$ upper limit, corresponding to a $3\sigma$ upper limit of $3.4\times10^{-4}$\,cts\,s$^{-1}$ in the 0.3--7\, keV band. Upper limits on fluxes can then be estimated using the \chandra\ PIMMS tool\footnote{\url{http://cxc.harvard.edu/toolkit/pimms.jsp}}. Assuming that the background emission can be modeled by a PL spectrum with $\Gamma = 1.5$ (typical value for the X-ray background), and adopting $N_{\rm H} = 2.7\times10^{21}$\,cm$^{-2}$ (molecular gas is present in front of AE Aur, but not towards the IR arc), the upper limits to the absorbed (unabsorbed) fluxes and luminosities are $6(\pm 8)\times10^{-15}$\,erg\,cm$^{-2}$\,s$^{-1}$ and $2(\pm 3)\times10^{29}$\,erg\,s$^{-1}$ (for $d=530$\,pc), respectively.

\subsection{Global Chandra/ACIS image analysis}
\label{chandra_global}

Without any preconceived ideas about the nature of the sources (even when there were obvious counterparts on MW images, which was rare; more below), we first ran our MW classification tool on all X-ray sources in the \chandra\ field (see Figure~\ref{chandra_merged}). The automated procedure produced eight classifications with a confidence level higher than 70\%, including three ``AGNs" and five ``stars". Note that the calculated confidence levels for each type do not include uncertainties associated with the X-ray flux determination, which can be substantial for faint X-ray sources; they also do not include the possibility of assigning false MW counterparts to the X-ray sources. Table~\ref{xray} shows the MW parameters (when available) for ObsID 19445. We do not list class designation and classification with confidence levels below 70\% because we do not deem those results accurate enough (the accuracy is lower usually due to the lack of MW counterparts). (Only the ten classified sources, together with CAX-72 and CAX-76, are displayed for brevity, while the entire table is available in electronic format.)

\begin{figure}
\begin{center}
\includegraphics[scale=0.55]{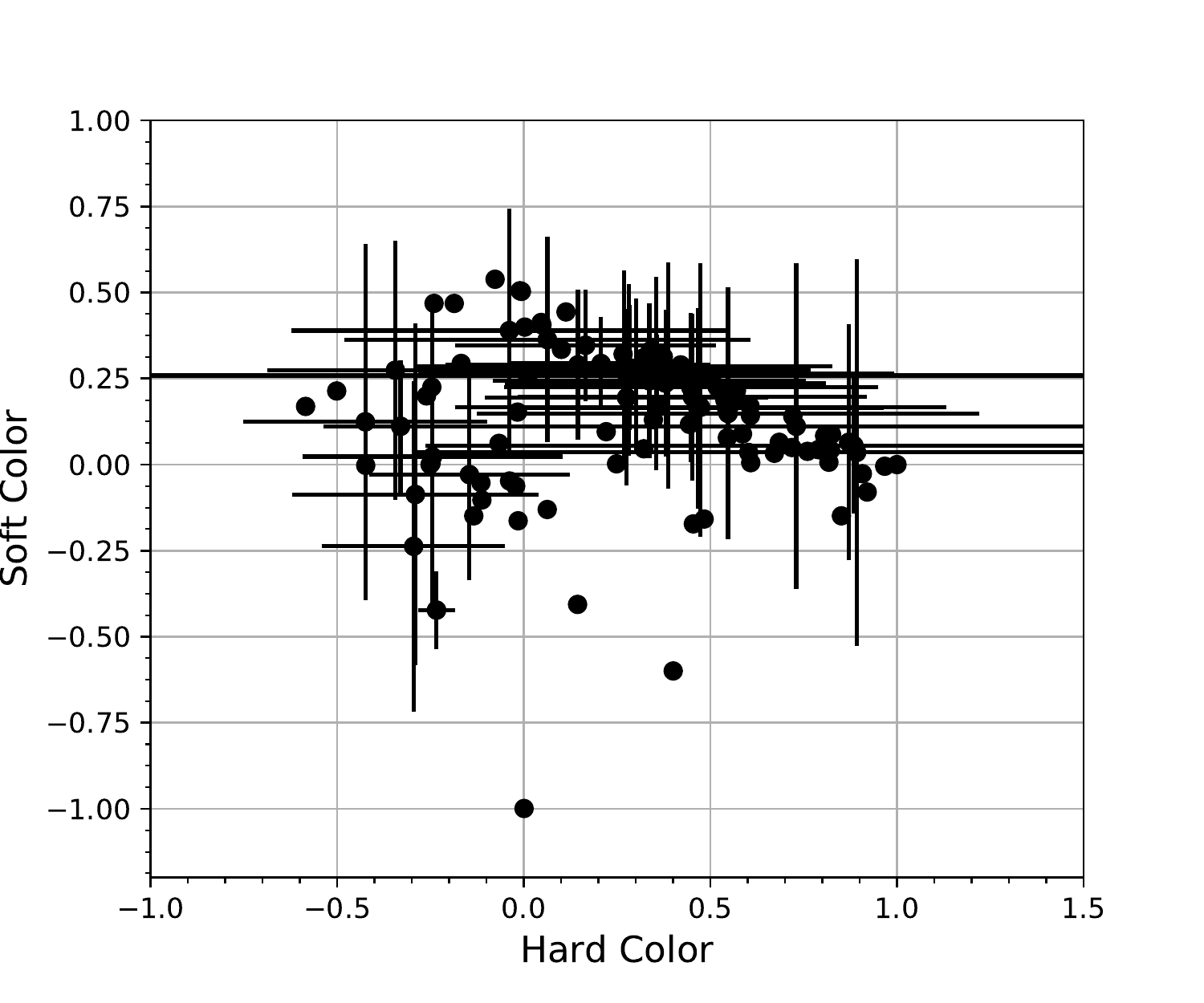}
\caption{Hardness ratio diagram for all sources based on all five observations. The two colors are created using the following equations {\it Soft Color = (M-S)/T} and {\it Hard Color = (H-M)/T}, where S, M, H and T are the counts in the soft (0.5--1.2\,keV), medium (1.2--2\,keV), hard (2--7\,keV) and broad (0.5--7\,keV) bands, respectively. Error bars are shown for sources with more than 40 counts in all observations.}
\label{hrd1}
\end{center}
\end{figure}

\begin{figure}
\begin{center}
\includegraphics[scale=0.55]{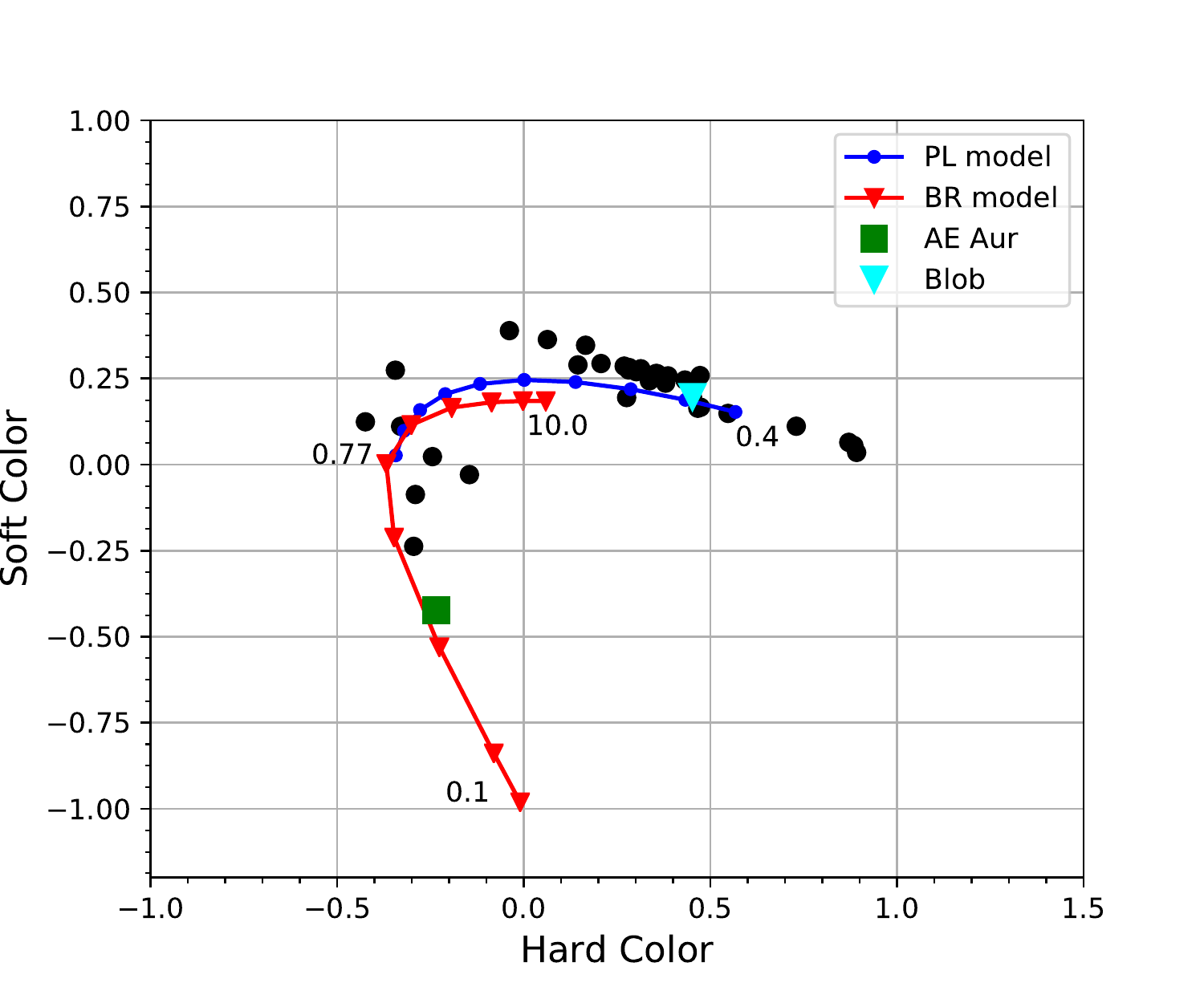}
\caption{Shown are absorbed PL (blue line and dots) with $\Gamma$ in the range of 0.4--4 ($\Delta \Gamma = 0.4$) and absorbed bremsstrahlung (red line and triangles) with $kT$ in the range of  0.1--10 (keV) in logarithmic steps ($\Delta log(kT)=0.33$). Both models have $N_{\rm H}=2.7\times10^{21}$\,cm$^{-2}$. For clarity, three of the model parameters (0.1, 0.77 and 10) are listed next to the symbols for the BR model, while only one (0.4) for the PL model. Only X-ray sources with more than 40 counts (in all observations) are plotted with black dots.}
\label{hrd2}
\end{center}
\end{figure}


\begin{deluxetable*}{lllrrccccc}
\tablecaption{X-ray sources (ObsID 19445)}
\tablehead{ \colhead{CAX \#} & \colhead{RA} & \colhead{DEC} & \colhead{Net\tablenotemark{a}} & \colhead{$F_X$\tablenotemark{b}} & \colhead{$I$\tablenotemark{c}} & \colhead{$J$\tablenotemark{d}} & \colhead{{\it W1}\tablenotemark{e}} & \colhead{Class} & \colhead{Probability} }
\startdata
12 & 78.9413882 & 34.2290756 & 3396.7 & $235\pm6.7$ & 10.89 & 13.752 & 10.726 & AGN & 84\% \\
19 & 78.9558607 & 34.3307404 & 11.8 & $0.5\pm0.3$ & 10.37 & 9.72 & 9.259 & STAR & 79\% \\
43 & 79.0092972 & 34.2252303 & 129.3 & $4.5\pm0.7$ &  & 10.766 & 10.244 & STAR & 72\% \\
51\tablenotemark{f} & 79.0315130 & 34.2805787   & 14.5  & $0.6\pm0.3$ &  & &    &    &    \\
72\tablenotemark{g} & 79.0756565 & 34.3122443 & 672.8 & $44.2\pm2.8$ & 5.82 & 5.342 & 5.309 & STAR & 82\% \\
74 & 79.0826617 & 34.3748928 & 96.9 & $4.1\pm0.7$ &  &  & 14.659 & AGN & 78\% \\
76\tablenotemark{h} & 79.0851935 & 34.3179972 & 17.7 & $1.1\pm0.4$ &  &  &  &  &  \\
85 & 79.1070493 & 34.3624738 & 18.4 & $0.4\pm0.2$ & 8.92 & 8.036 & 7.34 & STAR & 75\% \\
104 & 79.1933284 & 34.2357059 & 1226.2 & $65.6\pm3.1$ & 15.29 & 15.043 & 11.915 & AGN & 92\% \\
106 & 79.1969122 & 34.2458476 & 13.4 & $0.3\pm0.1$ & 15.46 & 14.248 & 13.281 & STAR & 71\%
\enddata 
\label{xray} 
\tablenotetext{a}{ Background subtracted net counts for ObsID 19445 only.}
\tablenotetext{b}{ X-ray flux in the 0.5--7\,keV band in units of $10^{-14}$\,erg\,cm$^{-2}$\,s$^{-1}$ for ObsID 19445.}
\tablenotetext{c}{ {\it I} magnitude from USNO-B1.}
\tablenotetext{d}{ {\it J} magnitude from 2MASS.}
\tablenotetext{e} {{\it W1} magnitude from {\it WISE}.}
\tablenotetext{f}{ This source is unidentified, but has a radio counterpart, EVS-2. See Table 3.}
\tablenotetext{g}{ This source is AE Aur.}
\tablenotetext{h}{ This source is the ``LS blob''.}
\tablecomments{ Only sources with confident classification and the ``LS blob'' are listed here. The available multi-wavelength parameters are used to classify these sources with our machine-learning pipeline (see the Appendix in \citealt{Hare2016} for details), which produces a classification (``Class'') and corresponding classification confidence (``Probability''). This table is available in its entirety in electronic format.}
\label{t2}
\end{deluxetable*}

To help with the investigation of the nature of the X-ray sources, we also made hardness ratio diagrams (see Figures~\ref{hrd1} and \ref{hrd2}) constructed as follows: {\it Soft Color = (M-S)/T} and {\it Hard Color = (H-M)/T}, where S, M, H and T are the counts in all five observations in the soft (0.5--1.2\,keV), medium (1.2--2\,keV), hard (2--7\,keV) and broad (0.5--7\,keV) bands, respectively. Figure~\ref{hrd2} shows two absorbed models: PL with $\Gamma$ in the [0.4, 4] interval with a step of $\Delta \Gamma = 0.4$, and bremsstrahlung with $kT$ in the [0.1, 10.0] (keV) interval in logarithmic scale with $\Delta log(kT)=0.33$. Both models have $N_{\rm H}=2.7\times10^{21}$\,cm$^{-2}$, the value obtained for AE Aur. These figures show that one cannot distinguish between PL models with $\Gamma \geq2$ and thermal models with $kT \ge 1$ (for the adopted absorption). We find in particular that CAX-72 (AE Aur) is consistent with a thermal model, typical of stellar sources, and that CAX-76 is definitely a hard source, with low $\Gamma$ and/or high absorption (more below).

\subsection{Evidence for radio emission from the bow shock?}

\begin{figure*}
\begin{center}
\includegraphics[scale=0.2]{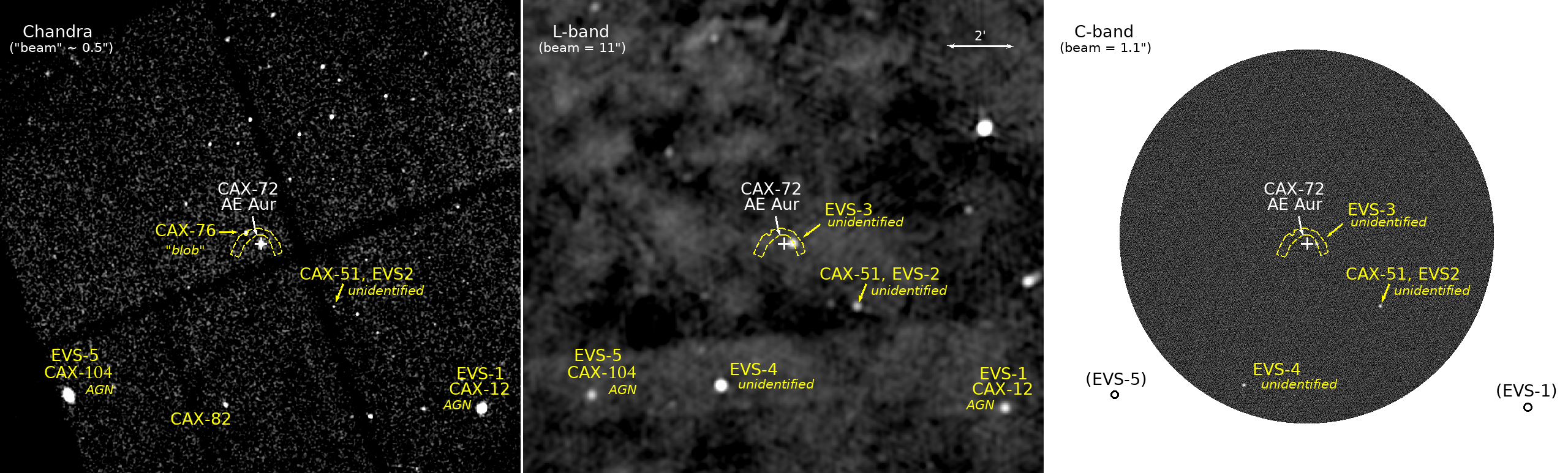}
\caption{X-ray (\emph{Chandra}) vs. Radio (EVLA) images centered on AE Aur ($16' \times 16'$). The EVLA C-band field and beam are much smaller than for the L-band because of the different EVLA configurations (C and B respectively). 
For comparison we have indicated the equivalent \emph{Chandra} ``beam'' of $\sim 0\farcs5$ (which is smaller along the axis; recall that the \emph{Chandra} PSF is not Gaussian). 
The radio sources discussed in the text are designated ``EVS-1" to ``EVS-5" (``EVS" standing for ``EVLA Source"). Identifications (and non-identifications) are indicated in italics. The position of AE Aur is indicated by a white cross, and the bow shock is outlined by the yellow dotted contour. The faint X-ray LS ``blob" is also indicated on the \chandra\ image. Further details are given in Table 3 and in the text (\S  3.5).}
\label{radio_blob}
\end{center}
\end{figure*}


\begin{deluxetable*}{lllcccc}
\tablecaption{Selected radio sources in the C- and L-band archival EVLA fields centered on AE Aur}
\tablehead{ \colhead{EVS-\#\tablenotemark{a}} & \colhead{RA} & \colhead{DEC} & \colhead{L flux (mJy)\tablenotemark{b}} & \colhead{C flux (mJy)\tablenotemark{c}} & \colhead{$\alpha$\tablenotemark{d}} & \colhead{Identification} }
\startdata
EVS-1 & 78.9412408  & 34.2294106    &  3.19 &    &   &  CAX-12 (AGN)  \\
EVS-2 & 79.0314167  & 34.2807639 & 1.47 & 0.93 & $- 0.36$ & CAX-51\tablenotemark{e} \\
EVS-3 & 79.0705606  & 34.3124611  & 3.24 & 0.81 & $- 1.08$ & unidentified \\
EVS-4 & 79.1143333  & 34.2407917  & 13.1 & 1.60 & $- 1.63$ & unidentified \\
EVS-5 & 79.1932599  & 34.2359209  &  2.27  &    &    &   CAX-104 (AGN)  \\
\enddata 
\label{evla}
\tablenotetext{a}{ ``EVS" = EVLA source (see Figure~\ref{radio_blob}).}
\tablenotetext{b}{ L band = tuned at 1.52 GHz (config. C: beamsize $11"$).}
\tablenotetext{c}{ C band = tuned at 5.5 GHz  (config. B: beamsize $1.1"$). Sources EVS-1 and EVS-5 lie outside the C-band FOV (see Figure~\ref{radio_blob}).}
\tablenotetext{d}{ $\alpha =$ spectral index from L frequency to C frequency.}
\tablenotetext{e}{ EVS-2 coincides with CAX-51 to better than $0.1"$, but has no other counterpart. Note that AE Aur itself is not detected. (See text for details.)}
\label{t3}
\end{deluxetable*}

The presence of energetic electrons associated with a bow shock might also be revealed by non-thermal radio emission, like that detected near the massive O star, BD+$43^{\circ}$3654 \citep{Benaglia2010}. The region of AE Aur has been observed with the EVLA in the C band in 2013 (B configuration; project 13B-212;  PI C. Peri) as well as in the L band in 2016 (C configuration; project 16A-152; PI C. Peri). The corresponding images obtained after performing a standard reduction of these archival data using the CASA\footnote{\url{https://casa.nrao.edu/}} software are displayed in Figure~\ref{radio_blob} together with the \emph{Chandra} X-ray image. Detailed inspection of these images near the position of AE Aur reveals no hint of extended emission at the position of the IR arc, in particular around the apex. Using the same procedure as described above for the X-ray emission (Section~3.3), we obtain $3\sigma$ upper limits of $1.1 \times 10^{-3}$ mJy/arcsec$^2$ and $1.7 \times 10^{-4}$ mJy/arcsec$^2$ on the brightness of the arc at 1.4 and 5 GHz, respectively, corresponding to upper limits for the flux emitted within the 1771 arcsec$^2$ area of 2 mJy and 0.4 mJy. A number of point sources are clearly detected in the L and C bands: their flux values together with the corresponding spectral indices ($\alpha$) and identifications are given in Table~3.

As shown in the composite \chandra-EVLA Figure~\ref{radio_blob}, two strong EVLA sources are detected in the L band (but not in the C band because of the reduced FOV): EVS-1 and EVS-5. Both show typical X-ray spectra of AGNs. Another EVLA source, EVS-3, is present near AE Aur but, with an RA offset of $-15''$, it is clearly distinct from the star. AE Aur (= CAX-72) is undetected in either band, which is consistent with the $3\sigma$ upper limit of 0.36 mJy previously obtained with the VLA at 6 cm (4860 MHz) by \cite{Bieging1989}. We also give flux values in Table 3 for two other sources detected in both bands, EVS-2 (= CAX-51) and EVS-4. Note that EVS-4 appears close to CAX-82 in Figure~\ref{radio_blob} but is not coincident with it. These three EVLA sources have no other counterpart. Given their steep spectral index, EVS-2 and EVS-3 could be distant non-thermal radio galaxies. 
 
 The region of AE Aur was also observed at 1.4 GHz and $45''$ resolution as part of the \emph{NRAO VLA Sky Survey} (NVSS; \citealt{Condon1998}), which allowed us to search for radio counterparts of our X-ray sources in the NVSS point source catalogue. Only two sources, CAX-12 (= EVS-1) and CAX-104 (= EVS-5), have NVSS counterparts, which further supports our findings that these are likely AGN (we checked that the fluxes given in the NVSS catalogue for the EVS-3 source and the radio counterparts to CAX-12 and CAX-104 are consistent with those measured in the EVLA L-band image). The spectra of both X-ray sources can be satisfactorily fitted using absorbed PL models with $N_{\rm H}\approx 10^{22}$\,cm$^{-2}$, in agreement with the constraints on $N_{\rm H}$ discussed in Sect.3.1 and the presence of some internal absorption. In Figure~\ref{agn_blob}, we display the spectra of CAX-72 (AE~Aur; thermal), CAX-12 (AGN; non-thermal) and  CAX-76 (the ``LS blob"). The spectrum of CAX-76 looks hard, even AGN-like, but the X-ray data alone (no MW counterpart was found) is not sufficient to fully characterize the true nature of this source.

\begin{figure}
\begin{center}
\includegraphics[scale=0.33]{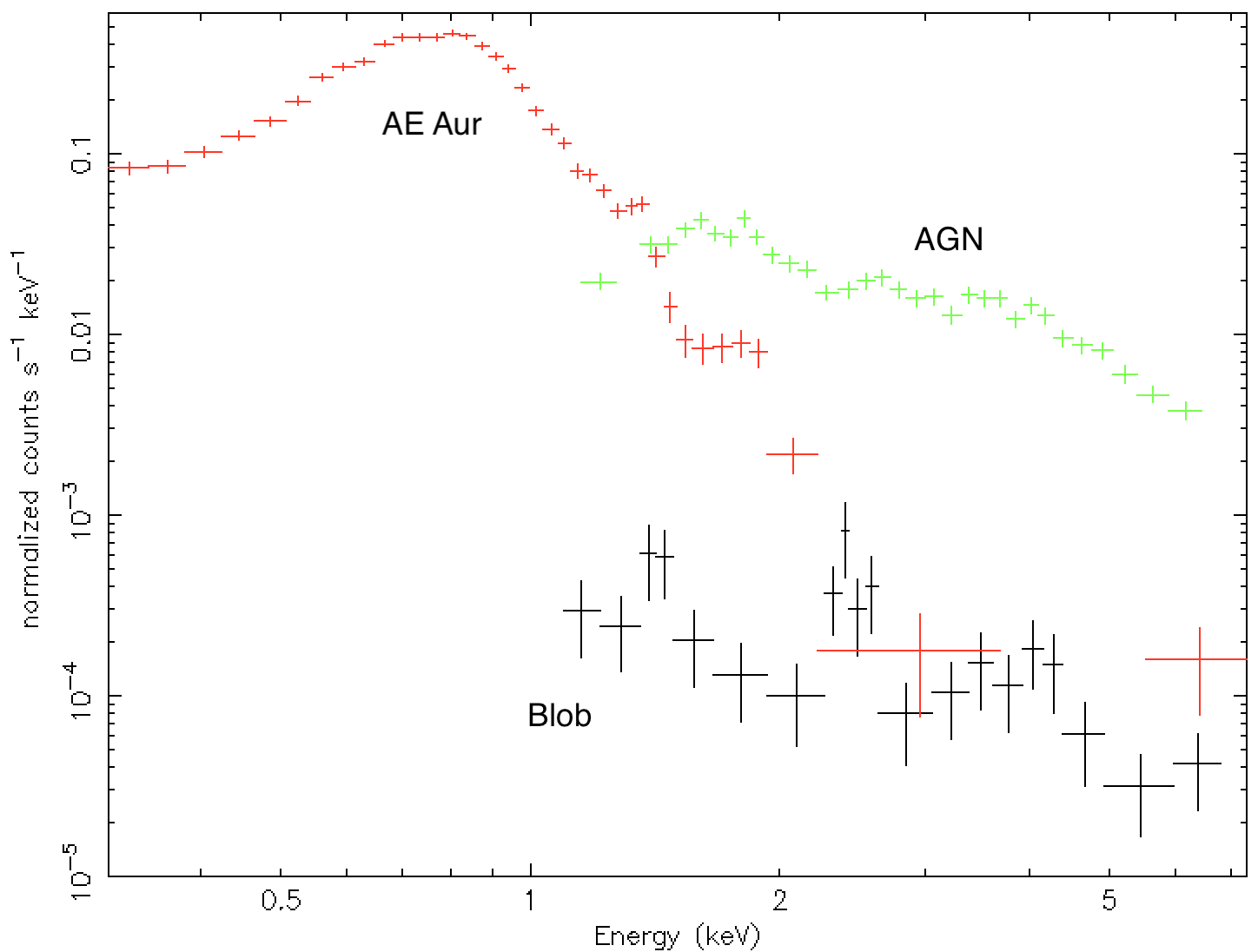}
\caption{X-ray spectra of three sources : ``blob" (= CAX-76; grouped by 5 counts per bin),  AE Aur (= CAX-72; 40 counts per bin), and one radio-detected AGN (EVS-1 = CAX-12; 100 counts per bin).}
\label{agn_blob}
\end{center}
\end{figure}

\section{Discussion}
\label{discussion}

\subsection{AE Aur in context}

As mentioned in the Introduction, AE Aur is one of the fastest runaway stars ($v_{\star}$ = 140 km~s$^{-1}$) showing a bow shock. This is the main reason why it was considered as a prime candidate to test particle acceleration by stellar bow shocks via their X-ray emission, but as we confirm in this paper, no diffuse, arc-shaped X-ray emission is seen when observed at high spatial resolution with \emph{Chandra}, nor is there any indication of radio emission being present from EVLA observations. 

However, other parameters, intrinsic to the star, may be important, from the point of view of momentum transfer rate from the shock to particles: the terminal wind velocity ($v_{\infty}$), and the mass-loss rate ($\dot{M}$). In addition, 2D hydrodynamic models by \citet{Green2019} show that the density of ISM and the orientation of the shock with respect to the observer are important. While, for massive stars, the terminal velocities are always comparable ($v_\infty \approx 1000-2000$ km~s$^{-1}$, which is of the same order as the escape velocity [$v_\infty = 2-3 \times v_{esc}$; see \citealt{Groenewegen1989}]), $\dot{M}$ may be very different, depending primarily on the spectral type (i.e., the UV radiation field), but even within the same spectral type (case of ``weak wind'' stars; \citealt{Shenar2017}). For AE Aur, $v_\infty = 1200$ km\,s$^{-1}$, and $\dot{M} = 2 \times 10^{-7} M_\odot$\,yr$^{-1}$.

In the above context, two more runaway bow shocks (from the list of \citealt{Peri2012}) have been recently studied in some detail, associated with one late O star (like AE Aur), and one early O star:  

\noindent
1) $\zeta$ Oph (O9.5Vnn; $d = 200$ pc; $v_{\star} = 24$ km\,s$^{-1}$): $v_\infty = 1500$ km\,s$^{-1}$, $\dot{M} = 2 \times 10^{-8} M_\odot/$\,yr$^{-1}$, one order of magnitude smaller than AE Aur, in spite of having the same spectral type;  and moving much slower;

\noindent
2) BD+43$^{\circ}$3654 (O4If; $d = 1.5$ kpc;  $v_{\star} = 14$ km\,s$^{-1}$): $v_\infty = 2300$ km\,s$^{-1}$, $\dot{M} = 6.5 \times 10^{-6} M_\odot/$\,yr$^{-1}$,  over one order of magnitude higher than AE Aur, and having a much earlier spectral type (higher mass, much more luminous); twice as slow as $\zeta$ Oph, and 10 times slower than AE Aur.

\subsection{Observational evidence for particle acceleration?}

The only direct indication so far of electron acceleration by a bow shock is the detection of resolved, extended radio synchrotron emission coinciding with that of BD+43$^{\circ}3654$, as seen in the IR \citep{Benaglia2010}, but not observed around AE Aur. However, the bow shocks of BD+43$^{\circ}$3654, and also of  $\zeta$ Oph (which does not show radio emission), were not detected in X-rays, respectively, by  \emph{Suzaku} \citep{Schulz2014} or \xmm\, and by \chandra\, or \suzaku\, \citep{Toala2016}. Extending the sample to six more runaway O stars in the \emph{XMM-Newton} archive did not produce more X-ray detections \citep{Toala2017}, nor did a study of a large sample of 60 IR-bright galactic stellar bow shocks by \chandra\ \citep{Binder2019}. On the other hand, the model for BD+43$^{\circ}$3654, based on the spectrum deduced from the radio emission, predicted that with \emph{Fermi} the IC dust \gray emission would be undetectable. Using a similar model it would be detectable for $\zeta$ Oph, because it is much closer (\citealt{VR2012}; in spite of its smaller $\dot{M}$ and  $v_{\star}$). However, this was not confirmed by a very sensitive \emph{Fermi} data analysis, where 27 bow shocks from \citet{Peri2012} were analyzed by \citet{Schulz2014}, with no detection and with upper limits $\sim 5$ times lower than predicted in a few cases \citep{Schulz2014}. Similarly, 32 bow shocks were analyzed using \hess\ data \citep{HESS2018}, with no detection. 

In this paper, we have obtained upper limits to the X-ray and radio emission from the AE Aur bow shock. In the next section, we show how these limits can be translated into theoretical constraints to understand whether standard particle acceleration model predictions are consistent with these systematic non-detections and only one positive detection, in the radio domain only. Recent developments in radiative models for stellar bow shocks themselves may also explain in part the current non-detections \citep{del Palacio2018}.

\section{Theoretical constraints on particle acceleration}

\subsection{Expected X-ray emission from the bow shock}

The available mechanical energy in bow shock systems is dominated by the kinetic luminosity of the stellar wind, which for AE Aur is equal to:

\begin{equation}
L_{w} = \frac{1}{2} \dot{M} v_{\infty}^2 \sim 10^{35} {\rm erg/s} ~ .
\end{equation}

A fraction $\eta_e$ of such energy can be converted into non-thermal electrons through DSA operating at the wind termination shock \citep{Benaglia2010}, and the accelerated electrons can in turn emit non-thermal radiation from radio frequencies up to the X-ray domain and possibly beyond (e.g. \citealt{del Palacio2018} and references therein). According to DSA theory, electrons are accelerated at the shock and injected in the system at a rate $Q_e(E_e) = Q_0 (E_e/m_e c^2)^{-2}$, where the normalisation constant $Q_0$ can be computed by imposing $\int {\rm d} E_e Q_e(E_e) E_e = \eta_e L_w$, which gives
 
\begin{equation}
Q_0 = \eta_e L_w/ (m_e c^2)^2
 \ln (E_{max}/m_e c^2) .
\end{equation}
\noindent
Here, $E_{max}$ is the maximum energy of the electrons accelerated at the shock and $m_e c^2$ the rest mass energy of the electron.

Electrons accelerated at the wind termination shock can produce non-thermal X-ray photons either as the result of IC scattering soft ambient photons, or via synchrotron emission in the magnetic field compressed (and possibly even amplified) at the shock. Let us consider first the IC scattering channel. The two most prominent photon targets to be considered are the radiation coming from the dust heated by the bow shock and that coming from the star. These two thermal radiation fields are characterized by temperatures of $T_d \sim 100$ K \citep{France2007} and $T_{\star} \sim 3.3 \times 10^4$ K \citep{Martins2015}, which in turn correspond to typical photon energies ($\sim 2.7 ~kT$, $k$ is the Boltzmann constant) equal to $\epsilon_d \sim 2 \times 10^{-2}$ eV and $\epsilon_{\star} \sim 8$ eV, respectively. In IC scattering, the energies of the electron $E_e$, of the ambient photon $\epsilon$, and of the upscattered photon $E_X$ are related as $E_X \sim \gamma^2 \epsilon$, where $\gamma = E_e/m_e c^2$ is the electron Lorentz factor. This implies that the electrons emitting IC photons in the X-ray band ($E_X \approx 1$ keV) are characterised by energies in the MeV domain ($E_e \approx 10-100$ MeV).

At such low energies, the radiative loss time of electrons is certainly much longer than the time $\tau_c$ needed to advect electrons away from the system (see e.g., Figure~1 in \citealt{Pereira2016}). Under these circumstances, the equilibrium spectrum of electrons is simply $N_e(E_e) \sim Q_e(E_e) \tau_c$, and the IC luminosity of the system $L_X$ can be estimated as: 

\begin{equation}
E_X^2 L_X(E_X) \sim N_e(E_e) P_{IC}(E_e) \frac{{\rm d} E_e}{{\rm d} E_X} E_X 
\end{equation}

\noindent where $P_{IC} \equiv {\rm d}E_e/{{\rm d} t}$ is the power emitted by an electron of energy $E_e$ in the form of IC photons. Note that $P_{IC}$ scales as the total energy density of soft ambient photons (dust emission plus radiation from the star). The IC flux observed at the Earth can then be expressed in a more convenient form by introducing the IC loss time $\tau_{IC} \equiv E_e/P_{IC}(E_e)$:

\begin{equation}
\label{eq:IC}
E_X^2 F_X(E_X) \sim \frac{\eta_e}{2} \left( \frac{L_w}{4 \pi d^2} \right) \frac{\tau_c}{\tau_{IC}(E_e)} \left[ \ln \left( \frac{E_{max}}{m_e c^2} \right) \right]^{-1} 
\end{equation}

\noindent where $d \sim 530$ pc is the distance to the AE Aur bow shock system.

The expected IC flux from AE Aur depends then on three poorly known quantities: the electron acceleration efficiency $\eta_e$, the ratio between the characteristic advection and IC time scales $\tau_c/\tau_{IC}$, and the maximum energy of the accelerated electrons $E_{max}$. In order to choose an appropriate value for $\eta_e$ we can proceed with an analogy with supernova remnant shocks, which are characterized by shock velocities and Mach numbers similar to those of wind termination shocks in bow shock systems. Supernova remnant shocks are believed to accelerate mainly cosmic ray {\it protons} (with an acceleration efficiency of $\approx 10$\%), and a number of observations (especially in the $\gamma$-ray domain) constrain the acceleration efficiency of electrons to values of the order of $\eta_e \approx 10^{-3}$ or less \citep{Cristofari2013,HESS2018}. Any realistic estimate of the ratio $\tau_c/\tau_{IC}$ should rely on detailed modeling, especially for what concerns the exact relative spatial distribution of accelerated electrons and soft ambient photons (coming both from the runaway star and from the dust heated by the bow shock). Even though this is not a straightforward task, several works seem to converge toward values of the order of $\tau_c/\tau_{IC} \approx 10^{-3} - 10^{-2}$ for the relevant electron energies  (see e.g., \citealt{Pereira2016,del Palacio2018} for a modeling of AE Aur and other bow shock systems). In particular, \citet{Pereira2016} claim that for the AE Aur system the energy density of the ambient radiation is largely dominated by dust emission, and found a value for the ratio $\tau_c/\tau_{IC}$ roughly equal to 10$^{-2}$. Finally, values of $E_{max}$ of the order of $\lesssim 1$ TeV have been estimated in \citet{Pereira2016}, implying that the logarithm in Eq. \ref{eq:IC} should be roughly of the order of $\approx 10$. Given these fiducial values we can now estimate the IC flux in the X-ray band as

\begin{multline}
\label{eq:B}
E_X^2 F_X(E_X) \lesssim 2 \times 10^{-15} \left( \frac{\eta_e}{10^{-3}} \right) \left( \frac{\tau_c/\tau_{IC}}{10^{-2}}\right) \\
\left[ \frac{\ln(E_{max}/m_e c^2)}{10} \right]^{-1} \rm erg/cm^2/s
\end{multline}

\noindent which is indeed an upper limit since it is based on the most optimistic value for $\eta_e$. 

Our estimate of the IC X-ray emission is consistent with the upper limit reported in Sec.~3.3, but is in disagreement with previous and more optimistic estimates reported in the literature (e.g., \citealt{Lopez-Santiago2012,Pereira2016,del Palacio2018}). The main reason for that is the different choice of the parameter $\eta_e$, which in these previous works was assumed to be equal to $\eta_e \approx 0.1$, which implicitly implies that wind termination shocks were assumed to be much more effective than SNR shocks in accelerating electrons. With this respect, our estimate is thus more conservative: we assumed that wind termination shocks behave as SNR ones.

To conclude, we stress that a similar reasoning can be adopted (referring to AE Aur) to interpret the radio upper limits. For the fiducial value of the magnetic field of $\sim 30 ~\mu$G, electrons of energy $\gtrsim$ 1 GeV emit synchrotron radiation in the GHz domain. The characteristic synchrotron cooling time $\tau_s$ for such electrons is of the order of several megayears. An estimate of the expected radio emission from the bow shock can then be obtained by substituting $\tau_{IC}$ with $\tau_s$ in Eq. 5 and by noticing that $\tau_c/\tau_s \approx 10^{-5}...10^{-4}$, which is about 2 orders of magnitude smaller than $\tau_c/\tau_{IC}$. However, this is almost exactly compensated by the fact that the radio upper limits are much more stringent than those derived from X-ray observations. Therefore we can conclude that also the non-detection of the bow shock of AE Aur in radio waves is consistent with the assumption that such shocks behave as SNR ones. 

Within this framework, the detection of radio synchrotron emission from the bow shock of the runaway star BD+43$^{\circ}$3654 might be tentatively explained by the very large kinetic energy characterising its stellar wind. For this system, estimates of the stellar mass loss rate range from $6.5 \times 10^{-6} ~ M_{\odot}$/yr \citep{Peri2012} up to $1.6 \times 10^{-4} ~ M_{\odot}$/yr \citep{Kobulnicky2010}, i.e. a factor of $\approx 30...800$ larger than that estimated for AE Aur. Also, the estimated velocity of the wind of BD+43$^{\circ}$3654 (2300 km/s) is roughly a factor of 2 larger than that estimated fro AE Aur \citep{Benaglia2010}. Therefore, the wind kinetic energy is $\approx 10^2 ... 3 \times 10^3$ times larger than that of AE Aur, which under certain conditions might explain the enhanced synchrotron radio emission from BD+43$^{\circ}$3654\footnote{We note that bow shocks associated with jets from young stars (e.g., Herbig-Haro objects) can accelerate particles and give rise to detectable radio synchrotron emission under certain conditions (e.g., \citealt{Padovani2015,Anglada2018}). Two recent examples are the HH80-81 complex of aligned knots \citep{RK2019}, or the jet from a massive star in the G035.02+0.35 star-forming region \citep{Sanna2019}. However, there are major differences with runaway bow shocks: in particular, in the cases mentioned, the magnetic field must be high ($B \sim 0.1-1$ mG), likely resulting from star-disk interactions at the base of the jets rather than being present in the ambient ISM; on the other hand, the successful VLA observations of HH objects are much more sensitive ($\sim 10 \mu$Jy/beam) than in the archival data for the AE Aur bow shock region ($\sim 1$ mJy/beam). (See also next subsection.)}.

\subsection{Maximum electron energy}

The estimate of $E_{max}$ deserves some further discussion. Although it has little impact on the estimate of the IC X-ray emission (it enters Eq.~\ref{eq:IC} as a logarithm), it might dramatically affect the estimate of the synchrotron X-ray emission. Electrons of energy $E_e$ gyrating around a magnetic field of strength $B$ radiate synchrotron photons of energy:

\begin{equation}
E_X^s \approx 0.1 \left( \frac{B}{100~\mu{\rm G}} \right) \left( \frac{E_e}{10~{\rm TeV}} \right)^2 ~ \rm keV
\end{equation}

\noindent so the question arises whether magnetic fields of the order of hundreds of $\mu$G could be found at wind termination shocks, and/or whether electrons can be accelerated there beyond 10\,TeV. 

\citet{del Palacio2018} noticed that such large values of the magnetic field at the shock are unlikely to be of stellar origin, but require some form of {\it in situ} amplification mechanism. One way to estimate the value of the amplified magnetic field is to persist with the analogy with SNR shocks. In SNRs the magnetic field can be significantly amplified at the shock due to plasma instabilities connected with the acceleration of cosmic rays (see e.g., \citealt{Bell2013}). X-ray observations of a number of SNR shocks showed that a small fraction $\xi_B$ (about few percent) of the shock ram pressure can be converted into magnetic field \citep{Volk2005}. If we assume that the position of the wind termination shock roughly coincides with the standoff radius $R_s$, i.e. the distance from the star where the ram pressure of the wind $\dot{M} v_{\infty}/4 \pi R_s^2$ balances that of the ambient medium $\varrho_{ISM} v_{\star}^2$ \citep{Wilkin1996}, the value of the magnetic field can be estimated as

\begin{equation}
\frac{B^2}{8 \pi}  = \xi_B \varrho_{ISM} v_{\star}^2 
\end{equation}

\noindent which gives:

\begin{equation}
B \approx 30 \left( \frac{\xi_B}{0.03} \right)^{1/2} \mu {\rm G} ~,
\end{equation}

\noindent where $\varrho_{ISM}$ is the mass density of the interstellar medium at the location of the bow shock. In order to obtain a conservative value for the magnetic field, we adopted a value of the ambient gas density of $\sim 3$ cm$^{-3}$, as reported in \citet{Peri2012}. A larger value of $\sim 20$ cm$^{-3}$ was found by \citet{Gratier2014}, and this would increase the estimate of $B$ by a moderate factor of $\sqrt{20/3} \sim 2.6$. One can see from Eq.~6 that for such a value of the magnetic field, the contribution of synchrotron emission to the X-ray flux would be relevant only if $E_{max}$ significantly exceeds 10\,TeV. Is such a value achievable at the wind termination shock? The model presented in \citet{Pereira2016} seems to suggest that this is not the case, and that the contribution from synchrotron radiation to the X-ray emission should be negligible.

It has been argued by \citet{Benaglia2010} that under certain conditions bow shock systems could indeed accelerate electrons to energies well beyond 1 TeV. An important consequence of this fact is that the IC emission from such systems could be potentially detectable in gamma rays both in the GeV and TeV domain, by instruments such as {\it Fermi} and the future \v{C}erenkov Telescope Array. Bow shock systems such has $\zeta$ Ophiuchi or BD+43$^{\circ}$3654 have been considered as potential $\gamma$-ray sources \citep{VR2012,del Palacio2018}. However, also in this case the estimates of the gamma-ray emission were based upon the assumption of a very efficient acceleration of electrons ($\eta_e \approx 0.1$). The adoption of a smaller value of $\eta_e$, as suggested by the analogy with SNR shocks, would make these systems too weak to be detected in $\gamma$-rays.

\section{Conclusions}
\label{conclusions}

A number of studies examined the possible acceleration of electrons at stellar runaway bow shocks (resulting from stellar winds colliding with the surrounding dense ISM), as revealed by X-rays from IC boosting of ambient IR shock dust emission. We revisited the case of the fast runaway, late O star AE Aur, obtaining high-angular resolution X-ray observations with \emph{Chandra}/ACIS observations. In short, confirming earlier findings based on \emph{XMM-Newton} and IR observations, we find pointlike X-ray emission at the location of the \emph{XMM-Newton} ``blob'' reported in the original paper by \citet{Lopez-Santiago2012}. However, no spatial coincidence with the IR signatures for the AE Aur bow shock occurs. 
 
In addition, to understand better the environment of AE Aur on the sky, both at small spatial scales (since dust emission is required for IC boosting to work), and at large scales (the area covered by the ACIS camera), we used CO and H~{\small I} data to estimate the gas column density in this region. These data allow a reliable estimate of the extinction, hence a reliable correction of the X-ray spectrum of the ``LS blob'' at low energies. Merging \emph{XMM-Newton} and \emph{Chandra} data, we find that, given its low count rate, it is not possible to find a good spectral fit, but the source is definitely hard and shows some similarities with the X-ray spectra of identified AGNs elsewhere in the field, suggesting that, although with no counterpart, it is likely an AGN as well. As a result, we confirm that, so far, no stellar runaway bow shock has been detected in X-rays, but for AE Aur we find an absorbed (unabsorbed) $3\sigma$ upper limit of $5.9(7.8) \times 10^{-15}$~erg~cm$^{-2}$~s$^{-1}$, which can be used to put stringent theoretical constraints on DSA models (see below).

By contrast, the only evidence so far for electron acceleration at stellar runaway shocks comes from the radio domain, with the detection of non-thermal emission from the early O star BD+43$^\circ$3654 associated with its bow shock. Therefore, we also investigated the possibility that the same situation might hold for AE Aur, by extracting data from the EVLA archives (L and C bands). The EVLA images show only point sources in the area. Neither AE Aur nor its bow shock are detected. An unidentified source is present in their close vicinity in both bands, but clearly distinct from them. Its two-point spectrum appears non-thermal; it may be also a weak background AGN. Calculated in a fashion similar to the \emph{Chandra} X-ray data, the EVLA upper limits ($3\sigma$) on the flux from the AE Aur shock are 2 mJy in the L band, and 0.4 mJy in the C band.

By comparison, the non-thermal radio emission associated with the BD+43$^{\circ}$3654 bow shock (slope $\alpha \sim -0.5$) stands out from thermal emission at the level of $\sim 100$ mJy, i.e., two orders of magnitude higher than the above upper limits for AE Aur. This is to be compared with its wind kinetic energy $L_w$, two to three orders of magnitude that of AE Aur: the stellar wind kinetic energy, not the runaway velocity, appears to be the dominant factor for the detection of their bow shocks, at least in the radio domain.

More generally, the above X-ray and radio upper limits can be used to put theoretical constraints on DSA models for stellar bow shocks. Two key ingredients are: ($i$) the fraction $\eta_e$ of the wind kinetic energy converted into non-thermal electrons via DSA, important mainly for X-ray generation (IC boosting of IR photons); ($ii$) the maximum energy $E_{max}$ of the accelerated electrons, important mainly for the synchrotron emission, via the ambient ISM magnetic field $B$; ($ii$) both are also important for $\gamma$-ray emission in the GeV-TeV range. Assuming that stellar wind shocks are analogous to SNR shocks, we have taken $\eta_e \sim 10^{-3}$ and $E_{max}$ significantly smaller than 10\,TeV, with $B \sim 30 \mu$G. With such values, we find that the X-ray, radio, and $\gamma$-ray emission for runaway stellar bow shocks are indeed undetectable at the current level of instrumental sensitivity, the case of BD+43$^{\circ}$3654 being the only exception so far. 

The future may lie in a few years, with new instruments like ESA's \emph{Athena} X-ray satellite, or the \emph{\v{C}erenkov Telescope Array} (CTA) from the ground in the 10 GeV-300 TeV domain. However, the angular resolution of these instruments is relatively poor. Our study has shown that unsuspected, non-thermal extragalactic sources like AGNs may lie spatially close to the targets, so that a careful study of their environment on the sky will be warranted before drawing any claim for detection of non-thermal emission from stellar runaway bow shocks.

\facility{{\it Chandra X-ray Observatory}, EVLA}

\acknowledgements

Support for this work was provided by the National Aeronautics and Space Administration through {\it Chandra} Award Number GO7-18131A issued by the Chandra X-ray Center, which is operated by the Smithsonian Astrophysical Observatory for and on behalf of the National Aeronautics Space Administration under contract NAS8-03060. This work has been financially supported by the Programme National Hautes Energies (PNHE). SG acknowledges support from the Observatoire de Paris under the program ``Action F\'ed\'eratrice CTA''. We thank Pierre Gratier for providing an appropriate version of the CO map to prepare our Figure~\ref{CO}.



\begin{thebibliography}

\bibitem[Abdo et al.(2010)]{Abdo2010} Abdo, A.~A., Ackermann, M., Ajello, M., et al.\ 2010, \apj, 712, 459 

\bibitem[Acciari et al.(2009)]{Acciari2009} Acciari, V.~A., Aliu, E., Arlen, T., et al.\ 2009, \apjl, 698, L133 

\bibitem[Anglada et al.(2018)]{Anglada2018} Anglada, G., Rodr{\'\i}guez, L.~F., \& Carrasco-Gonz{\'a}lez, C.\ 2018, \aapr, 26, 3

\bibitem[Arnaud(1996)]{Arnaud1996} Arnaud, K.~A.\ 1996, Astronomical Data Analysis Software and Systems V, 101, 17 

\bibitem[HI4PI Collaboration et al.(2016)]{Bekhti2016} HI4PI Collaboration, Ben Bekhti, N., Flöer, L., et al.\ 2016, \aap, 594, A116.

\bibitem[Bell et al.(2013)]{Bell2013} Bell, A.~R., Schure, K.~M., Reville, B., \& Giacinti, G.\ 2013, \mnras, 431, 415 

\bibitem[Benaglia et al.(2010)]{Benaglia2010} Benaglia, P., Romero, G.~E., Mart{\'{\i}}, J., Peri, C.~S., \& Araudo, A.~T.\ 2010, \aap, 517, L10 

\bibitem[Bieging et al.(1989)]{Bieging1989} Bieging, J.H., Abbott, D.C., \& Churchwell, E.B.\ 1989, \apj, 340, 518

\bibitem[Binder et al.(2019)]{Binder2019} Binder, B.~A., Behr, P., \& Povich, M.~S.\ 2019, \aj, 157, 176

\bibitem[Boiss{\'e} et al.(2005)]{Boisse2005} Boiss{\'e}, P., Le Petit, F., Rollinde, E., et al.\ 2005, \aap, 429, 509 

\bibitem[Cardillo et al.(2017)]{Cardillo2017} Cardillo, M., Ursi, A., Tavani, M., et al.\ 2017, GRB Coordinates Network, Circular Service, No.~21689, \#1 (2017), 21689, 1 

\bibitem[Christie et al.(2016)]{Christie2016} Christie, I.~M., Petropoulou, M., Mimica, P., \& Giannios, D.\ 2016, \mnras, 459, 2420 

\bibitem[Condon, \& Kaplan(1998)]{Condon1998} Condon, J.~J., \& Kaplan, D.~L.\ 1998, \apjs, 117, 361

\bibitem[Cristofari et al.(2013)]{Cristofari2013} Cristofari, P., Gabici, S., Casanova, S., Terrier, R., \& Parizot, E.\ 2013, \mnras, 434, 2748 

\bibitem[Cutri \& et al.(2012)]{Cutri2012} Cutri, R.~M., \& et al.\ 2012, VizieR Online Data Catalog, 2311,  

\bibitem[del Valle \& Romero(2012)]{VR2012} del Valle, M.~V., \& Romero, G.~E.\ 2012, \aap, 543, A56 

\bibitem[del Palacio et al.(2018)]{del Palacio2018} del Palacio, S., Bosch-Ramon, V., M{\"u}ller, A.~L., \& Romero, G.~E.\ 2018, \aap, 617, A13 

\bibitem[Drury(1983)]{Drury1983} Drury, L.~O.\ 1983, Reports on Progress in Physics, 46, 973 

\bibitem[Fischer et al.(2016)]{Fischer2016} Fischer, W.~J., Padgett, D.~L., Stapelfeldt, K.~L., \& Sewi{\l}o, M.\ 2016, \apj, 827, 96 

\bibitem[France et al.(2007)]{France2007} France, K., McCandliss, S.~R., \& Lupu, R.~E.\ 2007, \apj, 655, 920 

\bibitem[Freeman et al.(2002)]{Freeman2002} Freeman, P.~E., Kashyap, V., Rosner, R., \& Lamb, D.~Q.\ 2002, \apjs, 138, 185 

\bibitem[Gabici et al.(2009)]{Gabici2009} Gabici, S., Aharonian, F.~A., \& Casanova, S.\ 2009, \mnras, 396, 1629 

\bibitem[Gaisser et al.(2016)]{Gaisser2016} Gaisser, T., Engel, R., \& Resconi, E. (2016). \emph{Cosmic Rays and Particle Physics}. Cambridge: Cambridge University Press. doi:10.1017/CBO9781139192194

\bibitem[Gratier et al.(2014)]{Gratier2014} Gratier, P., Pety, J., Boiss{\'e}, P., et al.\ 2014, \aap, 570, A71 

\bibitem[Green et al.(2019)]{Green2019} Green, S., Mackey, J., Haworth, T.~J., Gvaramadze, V.~V., \& Duffy, P.\ 2019, \aap, 625, A4 

\bibitem[Groenewegen et al.(1989)]{Groenewegen1989} Groenewegen, M.~A.~T., Lamers, H.~J.~G.~L.~M., \& Pauldrach, A.~W.~A.\ 1989, \aap, 221, 78 

\bibitem[Hare et al.(2016)]{Hare2016} Hare, J., Rangelov, B., Sonbas, E., Kargaltsev, O., \& Volkov, I.\ 2016, \apj, 816, 52 

\bibitem[H.~E.~S.~S.~Collaboration et al.(2018)]{HESS2018} H.~E.~S.~S.~Collaboration, Abdalla, H., Abramowski, A., et al.\ 2018, \aap, 612, A3 


\bibitem[Hoogerwerf et al. (2001)]{Hoogerwerf2001} Hoogerwerf, R., de Bruijne, J. H. J., \&  de Zeeuw, P. T.\ 2001,
\aap, 365, 49

\bibitem[Ishibashi \& Courvoisier(2010)]{IC2010} Ishibashi, W., \& Courvoisier, T.~J.-L.\ 2010, \aap, 512, A58 

\bibitem[Katagiri et al.(2016)]{Katagiri2016} Katagiri, H., Sugiyama, S., Ackermann, M., et al.\ 2016, \apj, 831, 106 

\bibitem[Katsuta et al.(2012)]{Katsuta2012} Katsuta, J., Uchiyama, Y., Tanaka, T., et al.\ 2012, \apj, 752, 135 

\bibitem[Kim et al.(2007)]{Kim2007} Kim, M., Kim, D.-W., Wilkes, B.~J., et al.\ 2007, \apjs, 169, 401 

\bibitem[Kobulnicky et al.(2010)]{Kobulnicky2010} Kobulnicky, H.~A., Gilbert, I.~J., \& Kiminki, D.~C.\ 2010, \apj, 710, 549 

\bibitem[L{\'o}pez-Santiago et al.(2012)]{Lopez-Santiago2012} L{\'o}pez-Santiago, J., Miceli, M., del Valle, M.~V., et al.\ 2012, \apjl, 757, L6  (LS12) 

\bibitem[Malkov et al.(2011)]{Malkov2011} Malkov, M.~A., Diamond, P.~H., \& Sagdeev, R.~Z.\ 2011, Nature Communications, 2, 194 

\bibitem[Martins et al.(2015)]{Martins2015} Martins, F., Herv{\'e}, A., Bouret, J.-C., et al.\ 2015, \aap, 575, A34 

\bibitem[Monet et al.(2003)]{Monet2003} Monet, D.~G., Levine, S.~E., Canzian, B., et al.\ 2003, \aj, 125, 984 

\bibitem[Padovani et al.(2015)]{Padovani2015} Padovani, M., Hennebelle, P., Marcowith, A., et al.\ 2015, \aap, 582, L13

\bibitem[Pereira et al.(2016)]{Pereira2016} Pereira, V., L{\'o}pez-Santiago, J., Miceli, M., Bonito, R., \& de Castro, E.\ 2016, \aap, 588, A36 

\bibitem[Peri et al.(2012)]{Peri2012} Peri, C.~S., Benaglia, P., Brookes, D.~P., Stevens, I.~R., \& Isequilla, N.~L.\ 2012, \aap, 538, A108 

\bibitem[Peri et al.(2015)]{Peri2015} Peri, C.~S., Benaglia, P., \& Isequilla, N.~L.\ 2015, \aap, 578, A45 

\bibitem[Petruk et al.(2017)]{Petruk2017} Petruk, O., Orlando, S., Miceli, M., \& Bocchino, F.\ 2017, \aap, 605, A110 

\bibitem[Rodr{\'\i}guez-Kamenetzky et al.(2019)]{RK2019} Rodr{\'\i}guez-Kamenetzky, A., Carrasco-Gonz{\'a}lez, C., Gonz{\'a}lez-Mart{\'\i}n, O., et al.\ 2019, \mnras, 482, 4687

\bibitem[Sanna et al.(2019)]{Sanna2019} Sanna, A., Moscadelli, L., Goddi, C., et al.\ 2019, \aap, 623, L3

\bibitem[Schulz et al.(2014)]{Schulz2014} Schulz, A., Ackermann, M., Buehler, R., Mayer, M., \& Klepser, S.\ 2014, \aap, 565, A95 

\bibitem[Shenar et al.(2017)]{Shenar2017} Shenar, T., Oskinova, L.~M., J{\"a}rvinen, S.~P., et al.\ 2017, \aap, 606, A91 

\bibitem[Skrutskie et al.(2006)]{Skrutskie2006} Skrutskie, M.~F., Cutri, R.~M., Stiening, R., et al.\ 2006, \aj, 131, 1163 

\bibitem[Strong et al.(2007)]{Strong2007} Strong, A.~W., Moskalenko, I.~V., \& Ptuskin, V.~S.\ 2007, Annual Review of Nuclear and Particle Science, 57, 285 

\bibitem[Tavani et al.(2010)]{Tavani2010} Tavani, M., Giuliani, A., Chen, A.~W., et al.\ 2010, \apjl, 710, L151 

\bibitem[Tetzlaff et al.(2011)]{Tetzlaff2011} Tetzlaff, N., Neuh{\"a}user, R., \& Hohle, M.~M.\ 2011, \mnras, 410, 190 

\bibitem[Toal{\'a} et al.(2016)]{Toala2016} Toal{\'a}, J.~A., Oskinova, L.~M., Gonz{\'a}lez-Gal{\'a}n, A., et al.\ 2016, \apj, 821, 79 

\bibitem[Toal{\'a} et al.(2017)]{Toala2017} Toal{\'a}, J.~A., Oskinova, L.~M., \& Ignace, R.\ 2017, \apjl, 838, L19 

\bibitem[Torres et al.(2010)]{Torres2010} Torres, D.~F., Marrero, A.~Y.~R., \& de Cea Del Pozo, E.\ 2010, \mnras, 408, 1257 

\bibitem[Uchiyama et al.(2010)]{Uchiyama2010} Uchiyama, Y., Blandford, R.~D., Funk, S., Tajima, H., \& Tanaka, T.\ 2010, \apjl, 723, L122 

\bibitem[van Buren \& McCray(1988)]{BM1988} van Buren, D., \& McCray, R.\ 1988, \apjl, 329, L93 

\bibitem[V{\"o}lk et al.(2005)]{Volk2005} V{\"o}lk, H.~J., Berezhko, E.~G., \& Ksenofontov, L.~T.\ 2005, \aap, 433, 229 

\bibitem[Wilkin(1996)]{Wilkin1996} Wilkin, F.~P.\ 1996, \apjl, 459, L31 

\bibitem[Wright et al.(2010)]{Wright2010} Wright, E.~L., Eisenhardt, P.~R.~M., Mainzer, A.~K., et al.\ 2010, \aj, 140, 1868 



\end{thebibliography}
\end{document}